\documentstyle[amssymb,aps,multicol,prl]{revtex}

\draft

\begin{document}

\title{Temperature variations of the disorder-induced vortex-lattice melting landscape}

\author{A. Soibel$^1$, Y. Myasoedov$^1$, M. L. Rappaport$^1$, T. Tamegai$^{2,3}$, S. S. Banerjee$^1$,  and E. Zeldov$^1$}

\address{$^1$Department of Condensed Matter Physics,
The Weizmann Institute of Science, Rehovot 76100, Israel}
\address{$^2$Department of Applied Physics, The University of Tokyo, Hongo, Bunkyo-ku, Tokyo 113-8656, Japan }
\address{$^3$CREST, Japan Science and Technology Corporation (JST), Japan }
\date{\today}
\maketitle

\begin{abstract}
Differential magneto-optical imaging of the vortex-lattice melting
process in Bi$_2$Sr$_2$CaCu$_2$O$_8$ crystals reveals unexpected
effects of quenched disorder on the broadening of the first-order
phase transition. The melting patterns show that the
disorder-induced melting landscape $T_m(H,\textbf{r})$ is not
fixed, but rather changes dramatically with varying field and temperature
along the melting line. The changes in both the scale and shape of the
landscape are found to result from the competing contributions of different
types of quenched disorder which have opposite effects on the
local melting transition.

\end{abstract}

\pacs{PACS numbers: 74.60.Ec, 74.60.Ge, 74.60.Jg, 74.72.Hs}

\begin{multicols}{2}

In recent years, phase transitions in the vortex-matter have been
the subject of numerous theoretical and experimental
investigations \cite{Gianniu}. In particular, significant progress
has been achieved in experimental identification of the factors
that determine the mean-field location of the first-order melting
transition on the $H-T$ phase diagram. It has been shown that weak
point disorder shifts the melting transition to lower temperatures
while preserving its first-order nature \cite{Boris1,Paulius}.
Alternatively, presence of correlated disorder tends to shift the
transition to higher temperatures \cite{Boris2,Kwok}. Oxygen
doping, furthermore, changes the material anisotropy and
significantly alters the slope of the melting line
\cite{BorisPRL,Ooi}. In sharp contrast to the extensive knowledge
of the mean-field behavior, no information is currently available
on the factors that determine the {\em local} properties and
fluctuations of the phase transition. Since all the parameters
that determine the mean-field transition line $T_m(H)$ presumably
have considerable variance in their local values, the local
melting temperature, or melting landscape, $T_m(H, \textbf{r})$,
should exhibit significant local fluctuations, which generally
lead to rounding of a first-order transition \cite{Imry}.
Consequently, instead of undergoing a sharp melting transition in
the entire volume of the sample, the vortex-lattice may display a
non-trivial phase separation, the detailed mechanism of which is
unknown. Broadening mechanisms of the first-order transition are
of significant general interest in condensed matter physics \cite{Imry}.
However, their experimental study on an atomic scale is
formidable. The vortex system can therefore serve as an invaluable
tool for direct investigation of the local thermodynamic behavior at
phase transitions in the presence of quenched disorder.

In this Letter we present the first experimental study of
the effects of disorder on the local melting process at various
points along the first-order transition line. Variations in
quenched material disorder and anisotropy change the local vortex
potential and elastic moduli, thus creating a complicated melting
landscape $T_m(H, \textbf{r})$ in the sample. Using the recently
developed differential magneto-optical (MO) imaging \cite{Nature}
we have obtained a direct visualization of the melting landscape
in Bi$_2$Sr$_2$CaCu$_2$O$_8$ (BSCCO) crystals. In contrast to what
may be expected, the disorder-induced landscape is not constant,
but rather varies profoundly with changing temperature along the
melting line. It is found that the valleys of the landscape may
turn into peaks and that the characteristic lengthscale of
potential fluctuations increases by an order of magnitude with
decreasing temperature. These unexpected changes in both the scale
and shape of the melting landscape are shown to arise from
competing contributions of various types of disorder which
influence the melting transition differently.

The experiments were carried out on various BSCCO crystals
\cite{Ooi} ($T_c$ = 91 K) with typical dimensions  $ 1 \times 1
\times 0.05$ mm$^3$. In contrast to the conventional MO technique
\cite{MO}, in differential MO imaging the applied field $H_a$ is
modulated by a small amount $\delta H_a \ll H_a$ and the
corresponding differential signal is acquired using a CCD camera
and averaged over many cycles \cite{Nature}. This method improves
the sensitivity by about two orders of magnitude, yielding field
resolution of about 30 mG. In this work, we have used an
additional method, in which the sample temperature $T$ is
modulated by a small $\delta T \simeq 0.25$ K at a typical rate of
$10^{-3}$ Hz. The measurements were carried out by sweeping the
temperature or field both up and down through the transition with
either type of modulation. In the presented sample no irreversible
effects or hysteresis were observed throughout the melting
transition (compare with Fig. 4 in Ref \cite{Nature}).

The vortex-lattice melting is a first-order transition at which
the local field $B$ in the liquid phase increases discontinuously
by $\Delta B$ relative to the solid \cite{Zeldov}. On increasing
the temperature by $\delta T$, additional liquid droplets are
nucleated within the solid and the existing liquid domains expand.
In the differential image these new liquid regions appear as
bright areas in which the field is enhanced by $\Delta B \simeq
0.2 $ G. Figure 1 shows four examples of the differential melting
patterns at four different points along the melting line. In all
images liquid phase occupies about $50 \% $ of the crystal volume.
There are two striking observations in these images: one is that
the typical size and shape of the melting patterns change
dramatically with increasing field, and the other is that in most
of the sample area there is little correlation between the
patterns at low and high fields. At $H_a=20$ Oe the liquid regions
form arc-like patterns in the vertical direction. At a lower
field, 10 Oe, the arcs fracture into microscopic droplets each
containing just a few tens of vortices. At higher fields, in
contrast, the patterns become much coarser (40 and 75 Oe images)
and gradually lose correlation with the low-field patterns. From
75 Oe to 300 Oe, which is the upper limit of our measurements, the
melting patterns show only little change with field (not shown).
The melting patterns differ significantly from sample to sample,
but in all the investigated crystals similar characteristic
changes of the patterns with temperature are found.

Figures 2a and 2b present the evolution of the lattice melting as
the temperature is raised at a
fixed $H_a$, the different colors indicating areas which
melt upon 0.25 K increments of $T$. Material disorder modifies the local
melting temperature, thus forming a complicated $T_m(H,
\textbf{r})$ landscape. Figures 2a and 2b can thus be viewed as
`topographical maps' of the melting landscape at $H_a$ of 20 and
75 Oe, respectively. The minima points and the valleys of the
landscape melt first (blue), whereas the peaks of the landscape
melt last (red). The two landscapes in Fig. 2 are substantially
different. In addition to the significant change in the
characteristic lengthscale and roughness of the landscape, there
are many regions in the sample that show qualitatively different
properties. For example, at 20 Oe the valley in the form of an arc
along the `O-O' dashed line has three long and narrow blue
segments, while at 75 Oe, the blue minima have a form of rather
circular spots. Also, at 75 Oe in the lower-right corner of the
sample, to the right of the `O-O' valley, a number of blue and
cyan minima are visible. At 20 Oe, on the other hand, this region
is rather `elevated', characterized by green and yellow colors. In
the top part of the sample a yellow `ridge' is clearly visible
along the `P-P' line in Fig. 2b, whereas in Fig. 2a this ridge is
absent. At 75 Oe, on the right-hand-side of the ridge, there is an
extended peak (orange), whereas in the corresponding region at 20
Oe, blue and green valleys are seen. Also, importantly, the width
of the transition or the valley-to-peak height, changes
significantly. At 20 Oe the entire sample melts within about 1 K,
whereas at 75 Oe the melting process spans almost twice this
range.

For a more quantitative analysis we have inspected the melting
behavior at several points, e.g., points A and B in Fig. 2. The
melting lines $T_m(H)$ at these two points are shown in Fig. 3a.
There is a systematic divergence of the two melting lines that is
seen most clearly in Fig. 3b in the form of their difference
$T_{mB}-T_{mA}$ versus the mean-field $T_m$. Close to $T_c$, point
B melts 0.5 K below point A whereas at lower temperatures the
behavior is inverted and point B melts about 2 K above point A.
The two melting lines intersect at about 85 K. There are numerous
other points in the sample that show intersecting melting lines,
which means that regions which are valleys of the landscape at low
fields may turn into peaks at high fields and vice versa. Figure
3c shows the width of the melting transition of the entire
sample, which is the first direct measurement of the degree of
global rounding of the first-order transition along the melting
line. At high temperatures the vortex lattice melts within about
1 K whereas at low temperatures the transition width reaches about
5 K. All these findings show that the melting properties change
significantly along the melting line.

There are two main factors that determine the observed melting
behavior: one is the disorder-induced melting landscape and the
other is the solid-liquid surface tension $\sigma $. We analyze
first the effect of surface tension. The minimum nucleation radius
$r$ of a liquid droplet is determined by the balance between the
free energy gain at the transition $\Delta F = (\partial
F_s/\partial H -
\partial F_l/ \partial H) h\pi r^2d = \Delta Bhr^2d/4$ and the
energy cost of the
interface creation $2\pi \sigma rd$, which results in $r = 8\pi
\sigma / \Delta Bh$ \cite{Nature}. Here $h = H-H_m(T)$ is the
degree of superheating, $d$ is the sample thickness, and the
difference in the derivatives of the free energies in the solid
and liquid phases is given by $\partial F_s/\partial H -
\partial F_l/ \partial H=\Delta B/4 \pi$. The
surface tension can be estimated \cite{Nature} as $\sigma \simeq
\eta a_0H_m\Delta B/4\pi$, where $ \eta $ is a numerical
prefactor, $a_0 \simeq (\phi_0/H)^{1/2}$ is the intervortex
spacing, and $\phi_0$ is the flux quantum. $\sigma $ is thus
expected to grow as ${H_m}^{1/2}$ and hence the size of the
nucleating liquid droplets should increase accordingly as $ r
\simeq 2 \eta (\phi_0 H_m)^{1/2}/h $. Thus the increase with field
of the characteristic size of the liquid patterns in Figs. 1 and 2
could possibly be attributed to the increase in the surface
tension. However, the above estimate of the contribution of the
surface tension was carried out without disorder. In the presence
of disorder the variations in the nucleation radius due to a
change in $ \sigma $ are expected to be reduced. In contrast,
inspection of the 10 and 75 Oe patterns in Fig. 1, for example,
indicates that the increase in the characteristic droplet size is
substantially larger than the anticipated $(7.5)^{1/2}=2.74$.
Furthermore, the fact that above 75 Oe, up to 300 Oe, the scale
does not change with field is inconsistent with the surface
tension scenario. Yet the most compelling argument against the
dominant role of $ \sigma $ is the following. Surface tension
should act as a short-range filter that smears out the fine
structure of the melting landscape, but preserves the long-range
correlations. Analysis of the melting patterns shows that this is
not the case, and both the short and long range correlations are
lost with increasing $H_a$. In addition, it is highly unlikely
that surface tension can cause the local melting lines to cross as
in Fig. 3 and to change macroscopic valleys into peaks and vice
versa. Finally, a larger $ \sigma $ should sharpen the overall
melting transition by cutting off the tails of the
disorder-induced distribution function of the melting landscape
\cite{Imry}, in contrast to the observed broadening of the
transition with increasing $H_a$ (Fig. 3c). We therefore conclude
that the observed changes in the melting patterns are mainly
caused by changes in the disorder-induced potential landscape.
This conclusion is consistent with the previous assessment of a
very low solid-liquid surface tension \cite{Nature}.

We now address the possible mechanisms for the variations in the
potential landscape. For this it is convenient to parameterize the
local melting curves using the generic expression of the melting
transition \cite{BlatIvl}, $H_m(T,
\textbf{r})=H_0(\textbf{r})(1-T/T_c(\textbf{r}))^{\alpha
(\textbf{r})}$. We find that $\alpha (\textbf{r})$ depends weakly
on location, and hence set $\alpha (\textbf{r})= \alpha $ for
simplicity. Defining $T_c(\textbf{r})= T_c+ \Delta
T_c(\textbf{r})$ and $H_0(\textbf{r}) = H_0+\Delta
H_0(\textbf{r})$, we rewrite $H_m(T, \textbf{r})=H_m(T)+\Delta
H_m(T, \textbf{r})$, where $H_m(T)= H_0(1~-~T/T_c)^{\alpha}$ is
the mean-field melting line, and $\Delta H_m(T, \textbf{r})\simeq
[\Delta H_0(\textbf{r})+\alpha H_0 \Delta T_c(\textbf{r})/
(T_c~-~T)](1-T/T_c)^{\alpha}$ describes the disorder-induced
melting landscape. $\Delta H_m(T, \textbf{r})$ has two terms: the
first results from variations in the slope of the melting line
$\Delta H_0(\textbf{r})$ and the second from variations in the
local $T_c$. Due to the diverging $T_c-T $ denominator the second
term should become dominant near $T_c$, whereas the first term may
dominate at low temperatures. Analysis of the local melting lines
at various locations across the sample results in the following
parameters: $ H_0$ = 900 Oe, $\Delta H_0$ = 80 Oe, $T_c$ = 95 K,
$\Delta T_c$ = 0.7 K, and $\alpha$ = 1.6. Using these values we
find that the two terms of $\Delta H_m(T, \textbf{r})$ become
comparable at $T \simeq$ 82 K. This result explains a number of
observed features: Since at low temperatures $\Delta
H_0(\textbf{r})$ is the dominant term, the form of the landscape
becomes temperature and field independent, explaining the
invariance of the melting patterns above 75 Oe. However, the
amplitude of the landscape fluctuations increases as
$(1-T/T_c)^\alpha$ with decreasing temperature, thus explaining
the continuous broadening of the transition width in Fig. 3c. When
the two terms of $ \Delta H_m(T, \textbf{r})$ become comparable
above $\sim$ 82 K there is a gradual crossover to a new landscape.
As a result, the melting patterns change significantly and lose
their spatial correlations as seen in Fig. 1. At higher
temperatures the $\Delta T_c$ variations should eventually become
the dominant parameter which governs the shape and the scale of
the melting patterns. One can check the self-consistency of this
argument as follows. At high temperatures $\Delta T_c$ should also
modify the local critical field $H_{c1}(\textbf{r})$ and the
corresponding value of the local penetration field. Figure 2c
shows a high-sensitivity image of the initial field penetration
into the sample at $H_a$=2 Oe and $T$=89 K. A strong correlation
between the penetration form and the melting patterns at low
fields is readily visible. For a more accurate comparison Fig. 2d
presents a superposition of the penetration image with the melting
patterns at 20 Oe. The color in the image is given by the melting
contours, while the brightness is defined by the penetration
field. It is clearly seen that most of the macroscopic blue
regions of  liquid nucleation coincide with the bright areas where
the field penetrates first. In particular, the correspondence
between the arc structures of the penetration field and the
melting contours is remarkable. This correspondence indicates that
the melting propagation at low fields is indeed governed by the
local variations in $T_c$. It is interesting to note that a
comparison between the penetration image and the melting contours
at $H_a$ = 75 Oe results in a surprisingly large anti-correlation
behavior: The regions into which the field penetrates first are
often the last ones to melt. Such anti-correlation is clearly
seen, for example, for the `P-P' strip which is bright in Fig. 2c
but is mainly yellow and orange in Fig. 2b. The anti-correlation
behavior causes the observed crossing of the local melting lines
as in Fig. 3 and results in the variable melting landscape.

It is interesting to understand which types of material disorder
contribute to the different trends of the melting landscape.
Variations in pinning by point disorder may locally destabilize
the lattice, resulting in landscape minima. Local correlated
disorder, in contrast, is expected to stabilize the lattice,
creating maxima points of the landscape. However, such pinning
disorder is not expected to produce appreciable $\Delta T_c$
variations, nor should it result in crossing of the local melting
lines. Variations in the oxygen stoichiometry or in cation ratio,
on the other hand, could give rise to both effects. Oxygen doping
modifies both the $T_c$ and the anisotropy. In overdoped domains
$T_c$ is suppressed, but the corresponding reduction in anisotropy
results in a larger slope of $H_m(T)$ which may in turn lead to
the crossing of the local melting lines. Our analysis of the
crystal growth conditions indicates that the arc structures in
Fig. 2 are related to a slight modulation in material composition
and to the curved form of the meniscus between the solid and liquid
phases of BSCCO during the floating zone crystallization process.
We have also carried out preliminary X-ray spectroscopy
microanalysis \cite{Tamegai} which indicates some correlation
between small variations in the Sr/Cu ratio and the melting
patterns. This finding further indicates that the complexity of
the melting is related to the sample disorder and inhomogeneities.
However, the specific influence of the disorder still needs
further investigation since the same variation in Sr
concentration, for example, can have different microscopic effects
depending on the exact site occupied by excess Sr.

We thank D.E. Feldman and V.B. Geshkenbein for valuable
discussions. This work was supported by the Israel Science
Foundation and Center of Excellence Program, by Minerva
Foundation, Germany, by the Ministry of Science, Israel, and by
the Grant-in-Aid for Scientific Research from the Ministry of
Education, Science, Sports and Culture, Japan. EZ acknowledges the
support by the Fundacion Antorchas - WIS program.

%\newpage

%\newpage %%%% add for short

\ FIGURE CAPTIONS

Fig. 1. Differential MO images of the vortex-lattice melting in
BSCCO crystal at four points along the melting line: $H_m(T)$ = 10
Oe (89.5 K), 20 Oe (86.75 K), 40 Oe (82.0 K), and 75 Oe (75.25 K).
The differential images are obtained by subtracting the image at
$T-\delta T/2$ from the image at $T+\delta T/2$, with $ \delta T =
0.25 K$.

Fig. 2.  The contours of the melting propagation at $H_a$  = 20 Oe
(a)  and 75 Oe (b). The color code indicates the expansion of the
liquid domains as the temperature is increased in 0.25 K steps.
The onset of melting is at $T_m^{on} = 86.25 $ K in (a) and at 74.25
K in (b). (c) Differential magneto-optical image of the magnetic
field penetration at $H_a$ = 2 Oe, $T$ = 89 K, $\delta H_a$ = 1
Oe. (d) Superposition of (a) and (c).

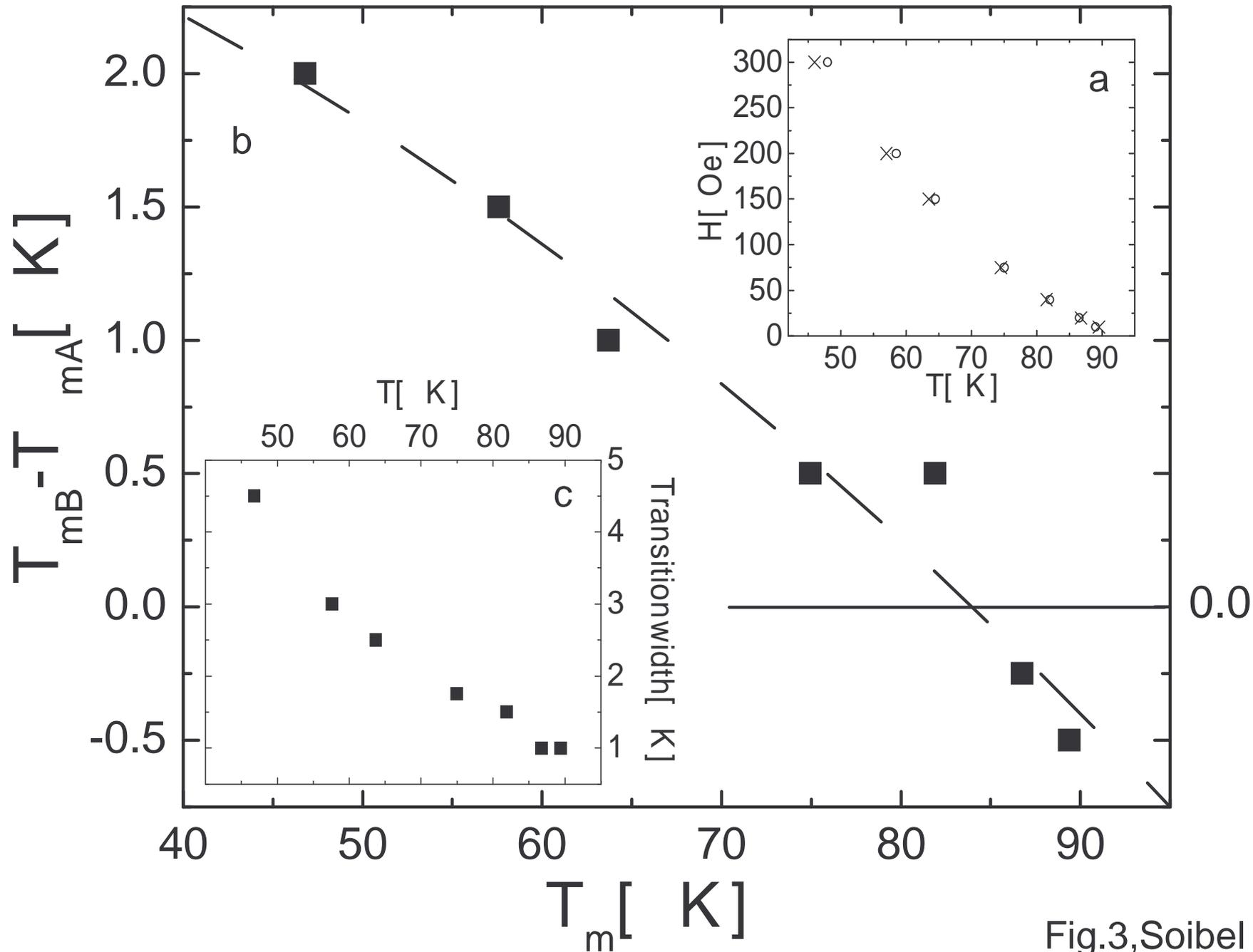
Fig. 3. (a) Local melting lines at points A $(\times)$ and B
$(\circ)$ indicated in Fig. 2. (b) Difference of the melting
temperatures at points B and A vs. the average melting
temperature. (c) The full width of the melting transition vs.
temperature.

\end{multicols}
%%%% add for long


\begin{references}

\bibitem{Gianniu}  G. Blatter {\it et al.}, Rev. Mod. Phys. {\bf 66}, 1125 (1994).

\bibitem{Boris1} B. Khaykovich {\it et al.}, Phys. Rev. B {\bf 56}, R517 (1997).

\bibitem{Paulius} L. M. Paulius {\it et al.}, Phys. Rev. B {\bf 61}, R11910 (2000).

\bibitem{Boris2} B. Khaykovich {\it et al.}, Phys. Rev. B {\bf 57}, R14088 (1998).

\bibitem{Kwok} W. K. Kwok {\it et al.}, Phys. Rev. Lett. {\bf 84}, 3706 (2000).

\bibitem{BorisPRL} B. Khaykovich {\it et al.}, Phys. Rev. Lett. {\bf 76}, 2555 (1996).

\bibitem{Ooi} S. Ooi  {\it et al.}, Physica C {\bf 302}, 339 (1998).

\bibitem{Imry}  Y.  Imry and M. Wortis,  Phys. Rev. B {\bf 19}, 3580 (1980).

\bibitem{Nature} A. Soibel  {\it et al.}, Nature {\bf 406}, 282 (2000).

\bibitem{MO} For a recent review see: A. A. Polyanskii {\it et al.}, NATO
Science Series {\bf 3/72}, 353 (Kluwer Academic Publishers, Dordrecht, 1999).

\bibitem{Zeldov} E. Zeldov  {\it et al.},  Nature {\bf 375}, 373 (1995).

\bibitem{BlatIvl}  G. Blatter and B. I. Ivlev, Phys. Rev. B {\bf 50}, 10272 (1994).

\bibitem{Tamegai} T. Tamegai, M. Yasugaki, K. Itaka, and M. Tokunaga,
Physica C {\bf 357-360}, 568 (2001).





\end{references}
\end{document}